\documentstyle[12pt]{article}
\begin{document}
\begin{titlepage}
\hfill{\vbox{\hbox{Published version of}\hbox{gr-qc/9408040}}}\bigskip
\begin{center}
\LARGE{\bf General Form of Thermodynamical Entropy for
Black Hole}\\ \vspace{1cm}
\large{{\bf A. Ghosh}\footnote{e-mail: amit@tnp.saha.ernet.in} and 
       {\bf P. Mitra}\footnote{e-mail: mitra@tnp.saha.ernet.in}}\\ \medskip
\normalsize{\em Saha Institute of Nuclear Physics, 1/AF Bidhannagar,\\
Calcutta 700 064, INDIA}\\ \medskip
\end{center}
\abstract{The  entropy  of  a  black hole can be different from a
quarter  of  the  area even at the semiclassical level. }  
\end{titlepage}

\def\wt {\widetilde}
\def\bb {\begin{equation}}
\def\ee {\end{equation}}
\def\pa {\partial}
\def\bea {\begin{eqnarray}}
\def\eea {\end{eqnarray}}

It  has long been known that black hole physics has a set of laws
parallel to  the  laws  of  thermodynamics \cite{BCH}.  By  virtue  of  this
parallelism,  the  area  of  the horizon of a black hole was
interpreted as its entropy \cite{Bek}.  After  the  discovery  of
Hawking  radiation  and  the  development of a semiclassical concept of
temperature for black holes, this analogy  became more well-defined
and  at  present a quarter of the area of the horizon is supposed
to be a quantitative measure of the entropy for non-extremal black
holes. No conclusive interpretation of
this  expression in terms of a counting of states has as yet been
found  in  spite  of serious attempts \cite{Uglum}. 
Reasonable progress has been made in the case of extremal black holes
\cite{Sen}, but there the entropy is proportional to the mass of the
black hole rather than the area \cite{GM}. In view of this departure
from the area law at the semiclassical level, it is natural to examine
the basis for the area formula and to see if deviations from the
standard expression   for   the   entropy  can  be  permitted  by
general principles also for non-extremal black holes.

For the  simplest  black  hole,  namely  the  one  discovered  by
Sch\-warzschild, the Hawking temperature is given by
\bb
T={1\over 8\pi M},\ee
where  $M$  is the mass of the black hole. Accordingly, the first
law of {\it thermodynamics} can be written as
\bb
dM=TdS=(8\pi M)^{-1}dS,\ee
which shows that the entropy must be $4\pi M^2$ upto an  additive
constant,  {\it  i.e.},  essentially  a  quarter  of the area. No
analogy is involved here, and the  standard  result  is  obtained
directly  from  thermodynamics.  There is no scope for ambiguity.
However, this is no longer the case if  we  go  on  to  more
complicated   black   holes.  We  shall  demonstrate  below  that
thermodynamics allows some freedom  in  the  expression  for  the
entropy  in the case of black holes depending on extra parameters
like  charge  or  angular momentum. The details will be worked out
in the case of the Kerr-Newman black hole.

This solution may be regarded as a rotating, charged extension of the
Schwarzschild black  hole.
There are three parameters: $\,M\,$, the mass of the
black hole, $Q$, its charge and $\,J\,$, the angular
momentum. There  are  two  horizons occurring at
\bb
r_{\pm}=M\pm\sqrt{M^2-{J^2\over M^2}-Q^2}.\ee
The area of the outer horizon is given by
\bb
A=4\pi (r_+^2+J^2M^{-2})\ee
and the Hawking temperature by
\bb
T={r_+-r_-\over 4\pi (r_+^2+J^2M^{-2})}.\ee
The first law of black hole physics \cite{BCH} takes the form 
\bb
Td(A/4)=dM-\Phi dQ-\Omega dJ\label{A},\ee
where the potential is given by
\bb
\Phi=\left.{\pa M\over\pa Q}\right|_A={Qr_+\over r_+^2+J^2M^{-2}}\ee
and the angular velocity by
\bb
\Omega=\left.{\pa M\over\pa J}\right|_A={JM^{-1}\over r_+^2+J^2M^{-2}}.\ee

The important point to note is that these derivatives are calculated
at constant {\it area}, whereas chemical potentials
are supposed to be calculated by differentiating
at constant {\it entropy}. It would be circular to try to argue that
the area is a measure of the entropy
by identifying these derivatives with  chemical potentials. {\it We shall
avoid this trap.}

Comparing (\ref{A}) with the first law of thermodynamics
\bb
TdS=dM-\wt\Phi dQ-\wt\Omega dJ,\ee
where $\wt\Phi,\wt\Omega$ are the unknown chemical potentials
corresponding to $Q,~J$ respectively, {\it i.e.,}
\bb
\wt\Phi=\left.{\pa M\over\pa Q}\right|_S,\ee
\bb
\wt\Omega=\left.{\pa M\over\pa J}\right|_S,\ee
one can write
\bb
d(S-{A\over 4})=-{\wt\Phi-\Phi\over T}dQ- {\wt\Omega-\Omega\over T}dJ.\ee
Since the left hand side is an exact differential, one must have
\bb
{\wt\Phi-\Phi\over T}={\pa F(J, Q)\over \pa Q}\ee 
and
\bb
{\wt\Omega-\Omega\over T}={\pa F(J, Q)\over \pa J}\ee
with $\,F(J, Q)\,$ undetermined. Consequently
\bb
S={A\over 4}-F(J, Q).\ee
{\it This is the general form of the entropy, and it involves an undetermined
function of the charges of the black hole.}

The function $F$ can be sought  to  be fixed from other arguments. In a  
functional integral approach, one may consider the partition function
\bb
Z=e^{iI_E}\ee
where $\,I_E\,$ is an effective action defined by evaluating the 
functional integral \cite{GH}. $\,Z\,$ must be interpreted as
the grand canonical partition function
\bb
Z=e^{-W/ T},\ee
where
\bb
W=M- TS-\wt\Phi Q-\wt\Omega J.\ee
In the stationary  phase approximation the functional integral is
taken  to  be  dominated  by  the  classical  configuration   and
$\,I_E\,$ is approximately equal to $i(M-\Phi Q)(2T)^{-1}$ \cite{GH}.
It follows  that
\bb
{M-TS-\wt\Phi Q-\wt\Omega J\over T}={1\over 2T}(M-\Phi Q).\ee
By using the generalized Smarr formula \cite{Smarr, BCH} 
\bb
M={TA\over 2}+\Phi Q+2\Omega J\label{Smarr}\ee
and  substituting the expressions for $S,~\wt\Phi$ and $\wt\Omega$, we
find
\bb
F=J{\pa F\over\pa J}+Q{\pa F\over\pa Q}.\ee
The general solution of this equation is a homogeneous function
of degree 1 (but not necessarily {\it linear}) in $J,~Q$. 

We started from the first law of black hole physics and showed that the first
law of thermodynamics allows the entropy to differ from $A/4$ by an arbitrary
function of the charges, {\it i.e.,} the electric charge and the angular
momentum in the Kerr-Newman case considered here. Thereafter, we took the
partition function in the leading semiclassical approximation into account and
found some restriction on the function $F$. One may wonder if further
restrictions can be imposed if the recently developed microcanonical
partition function \cite{york} is taken into consideration. A detailed
look at the analysis of that paper shows, however, that the restriction cannot 
be sharpened that way. This is essentially because the microcanonical action
introduced in \cite{york} is derived from the usual action involved in the
grand partition function of \cite{GH} and can only lead to equivalent
results. To be precise, the microcanonical action has been defined by
\bea
{\cal S}_m-{\cal S}&=&\int_{^3B}d^3x\sqrt\sigma(N\epsilon - Vj)\nonumber\\
&=&-{iM\over T}+{i\Omega J\over T}\eea
when $N,~V$ (see Eq. (3.9a) and Sec. VI of \cite{york}) are constant on $B$.
Now the potential $V$ related to $\Omega$
is not unique and can be modified in the way $\Omega$ of our paper is
replaced by $\wt\Omega$. This will introduce in the
microcanonical entropy the kind of ambiguity that we have identified above.
If a more direct and fundamental way could be found to introduce a 
microcanonical partition function, the ambiguity in the entropy could
perhaps be removed.

Our derivation obviously generalizes to the case of more parameters,
where the parameters are kept independent of one another and of
the mass {\it i.e.,} non-extremal black holes are considered. In short,
then, we have demonstrated  that such black  holes
do not have their entropies uniquely determined. 
The general form of the ambiguity involves
undetermined homogeneous functions of degree 1 in the parameters,
{\it  i.e.,}  charge, angular momentum and so on.

\end{document}